\documentclass[aps,prb,amsmath,amssymb,twocolumn,floatfix]{revtex4-1}
\usepackage{graphicx}% Include figure files
\usepackage{dcolumn}% Align table columns on decimal point
\usepackage{bm}% bold math

\begin{document}

\title{Comment on ``Discovery of slow magnetic fluctuations and critical slowing down in the pseudogap phase of YBa$_2$Cu$_3$O$_y$''}

\author{Jeff E. Sonier,$^{1,2}$}

\affiliation{$^1$Department of Physics, Simon Fraser University, Burnaby, British Columbia, Canada V5A 1S6 \\
$^2$Canadian Institute for Advanced Research, Toronto, Ontario, Canada M5G 1Z8}

\date{\today}
%%%%%%%%%%%%%%%%%%%%%%%%%%%%%%%%%%%%%%%%%%%%%%%%%%%%%%%
\begin{abstract}
A recent zero-field (ZF) and longitudinal-field (LF) muon spin relaxation ($\mu$SR) study of YBa$_2$Cu$_3$O$_y$ 
[Jian~Zhang {\it et al.}, arXiv:1709.06799] claims to have detected critical slowing down of magnetic
fluctuations near the pseudogap temperature $T^*$, and attribute this to the onset of
slow fluctuating domains of intra-unit-cell magnetic order. Here it is argued that the relaxation data 
displayed in this study are misleading due to an improper account of the nuclear dipole contribution
and a failure to acknowledge the occurrence of muon diffusion.
\end{abstract}

\maketitle
%%%%%%%%%%%%%%%%%%%%%%%%%%%%%%%%%%%%%%%%%%%%%%%
In a ZF or LF $\mu$SR experiment, the measured asymmetry spectrum $A(t)$ is proportional to the time evolution of the muon spin 
polarization $P_z(t)$ along the initial direction of the muon spin (defined to be the $z$ axis). Each positive muon ($\mu^+$) 
implanted in the sample senses a resultant field 
\begin{equation}
{\bf B}_{\mu} = {\bf B}_{\rm dip} + {\bf B}_{\rm con} + {\bf B}_{\rm LF} \, ,
\end{equation}
where ${\bf B}_{\rm dip}$ originates from magnetic dipole moments inside the sample,
${\bf B}_{\rm con}$ is the Fermi contact field in conductors generated by the net spin density of conduction
electrons in contact with the $\mu^+$, and ${\bf B}_{\rm LF}$ is the
external magnetic field that is applied in the $z$ direction in a LF experiment. In general, ${\bf B}_{\rm dip}$ consists of
contributions from both nuclear and electronic moments. The electronic dipolar field sensed by the $\mu^+$ may be 
static or fluctuating. Since the correlation times of the nuclear moments are typically much longer than the muon life 
time, the resultant nuclear dipolar field sensed by the $\mu^+$ is generally considered to be static.   

Zhang {\it et al.} make the implicit assumption that the nuclear dipole contribution to the ZF-$\mu$SR
signals of YBa$_2$Cu$_3$O$_{6.72}$ and YBa$_2$Cu$_3$O$_{6.95}$ is $T$-independent and well described by what 
is known as a static Gaussian Kubo-Toyabe (KT) relaxation function.\cite{Kubo:81} Neither of these assumptions 
are valid. The reasons why can be found in an earlier ZF-$\mu$SR study of YBa$_2$Cu$_3$O$_y$ (Ref.~\onlinecite{Sonier:02}).
There it was shown that the nuclear dipole contribution to the ZF-$\mu$SR signal is modified 
(1) below $T \! \sim \! 100$~K by the formation of charge-density-wave order (CDW) in the CuO chains, (2) an apparent
structural change near 60~K, and (3) above $T \! \sim \! 160$~K by muon diffusion.       

Zhang {\it et al.} fit their recorded ZF-$\mu$SR spectra for YBa$_2$Cu$_3$O$_{6.72}$ and YBa$_2$Cu$_3$O$_{6.95}$ with 
the following relaxation function
\begin{equation}
G_z(t) = G_{\rm KT}(\Delta, t) e^{- \lambda_{\rm ZF} t} \, ,
\label{eqn:GE}
\end{equation}
where 
\begin{equation}
G_{\rm KT}(\Delta, t) = \frac{1}{3} + \frac{2}{3} (1- \Delta^2 t^2)e^{- \frac{1}{2} \Delta^2 t^2} \, ,
\label{eqn:AsyZF}
\end{equation}
is the static Gaussian KT function. They assumed the relaxation rate $\Delta$ is $T$-independent and 
that $G_{\rm KT}(\Delta, t)$ fully accounts for the static nuclear dipole fields. The exponential relaxation rate 
$\lambda_{\rm ZF}$ is intended to account for static or fluctuating electronic fields. The value of $\Delta$ was 
determined from global fits of the the ZF-$\mu$SR spectra over the entire temperature range. 
Strictly speaking, the static Gaussian KT function pertains to a dense
system of randomly oriented frozen spins that produce an isotropic Gaussian distribution of dipolar field at the 
$\mu^+$ site. In actuality this situation is never realized in a real material, and the quadrupole interactions of the
the nuclear spins with the electric field gradients generated by the muon and other sources 
must be taken into acount.\cite{Holzschuh:84} 
Consequently, $G_{\rm KT}(\Delta, t)$ is a crude approximation of the nuclear dipole 
contribution. It serves as an adequate approximation when electronic dipoles dominate the relaxation of the
ZF-$\mu$SR signal. However, this is not the case in the experiments by Zhang {\it et al.},
where $\Delta/\gamma_\mu \! \approx \! 1.2$~G and $\lambda_{\rm ZF}/\gamma_\mu \! < \! 0.4$~G 
($\gamma_\mu/2 \pi \! = \! 135.54$~MHz/T is the muon gyromagnetic ratio). In the past strong deviations of 
the nuclear dipole contribution from $G_{\rm KT}(\Delta, t)$ were recognized and appropriately modeled 
in pure Cu,\cite{Celio:86} but also more recently in La$_{2-x}$Sr$_x$CuO$_4$.\cite{Huang:12} 
As shown in these studies, an accurate description of 
the magnetic dipolar coupling of the $\mu^+$ to the host nuclei requires calculations that include the electric
quadrupolar interaction of the nuclei with the local electric-field gradient (EFG). This is because
the quantization axis for the nuclear spins in ZF is oriented along the direction of the maximal local EFG.

\begin{figure}
\centering
\includegraphics[width=10.2cm]{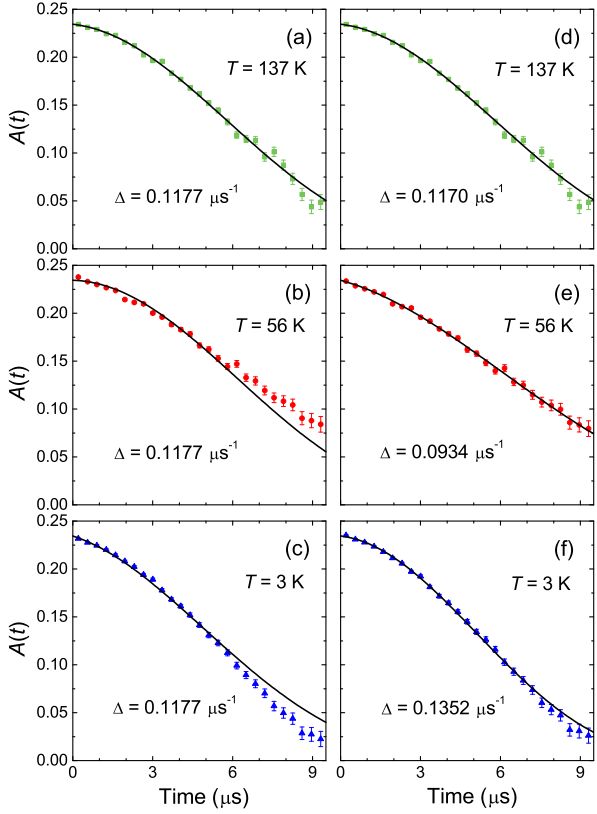}
\caption{(Color online) Fits of representative ZF-$\mu$SR spectra for YBa$_2$Cu$_3$O$_{6.985}$ assuming
(a)-(c) $\Delta$ is $T$-independent as in Ref.~\onlinecite{Zhang:17}, and (d)-(f) $\Delta$ varies with $T$
as in Ref.~\onlinecite{Sonier:02}.}
\label{fig1}
\end{figure}

The EFG at nuclei in an ionic crystal lattice is modified by the presence of the $\mu^+$ and in metallic systems by
its screening cloud of conduction electrons. In YBa$_2$Cu$_3$O$_y$ the local EFG is also modified by the onset of 
static CDW order in the CuO chains. Measurements of the nuclear quadrupole resonance (NQR) linewidth $\delta \nu_Q$ in 
YBa$_2$Cu$_3$O$_7$ show an abrupt increase in the local charge distribution around the Cu(1) chain sites and
Cu(2) plane sites across $T_c$.\cite{Grevin:00} Moreover, the variation of $\delta \nu_Q$ with $T$ at the Cu(2) site is
nonmonotonic, exhibiting a broad hump near 60~K and a dip near 40~K. The in-plane charge modulation is apparently
induced by the in-chain CDW correlations. The effect of the CuO chain CDW state on the ZF-$\mu$SR signal
was clearly demonstrated in Ref.~\onlinecite{Sonier:02}. There it was recognized that
Eq.~(\ref{eqn:GE}) can only be used to describe the variation of the ZF-$\mu$SR spectrum with temperature if
the assumption of $\Delta$ being $T$-independent is lifted. Figure~\ref{fig1}
shows a comparison of fits of the ZF-$\mu$SR spectra of YBa$_2$Cu$_3$O$_{6.985}$ with $\Delta$ as a $T$-independent parameter
to fits achieved with $\Delta$ free to vary with $T$. As in the study by Zhang {\it et al.}, the $T$-independent
value of $\Delta$ was obtained by a global fit of ZF-$\mu$SR spectra recorded over the full temperature range.
Note that the deviation of the fits with a $T$-independent $\Delta$ from the measured ZF-$\mu$SR signal becomes 
apparent beyond 6~$\mu$s, necessitating an analysis of the spectra to longer times.       

\begin{figure}
\centering
\includegraphics[width=8.5cm]{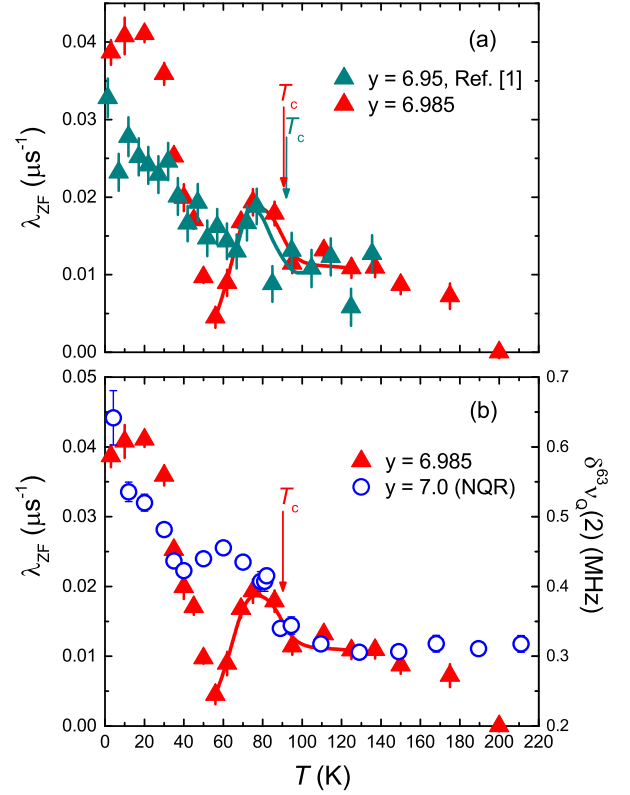}
\caption{(Color online) Results of fits to Eq.~(\ref{eqn:GE}) assuming $\Delta$ is $T$-independent. 
(a) Comparison of the temperature dependence of the exponential relaxation rate $\lambda$
in YBa$_2$Cu$_3$O$_{6.95}$ from Ref.~\onlinecite{Zhang:17} to $\lambda$ obtained from fits of the YBa$_2$Cu$_3$O$_{6.985}$
ZF-$\mu$SR spectra reported in Ref.~\onlinecite{Sonier:02}.
(b) Comparison of the temperature dependence of $\lambda$ in YBa$_2$Cu$_3$O$_{6.985}$ to the temperature dependence
of the Cu(2) NQR linewidth in YBa$_2$Cu$_3$O$_7$ reported in Ref.~\onlinecite{Grevin:00}.}
\label{fig2}
\end{figure}
Figure~\ref{fig2}(a) shows a comparison of the temperature dependence of $\lambda_{\rm ZF}$ in YBa$_2$Cu$_3$O$_{6.95}$ from 
Ref.~\onlinecite{Zhang:17} to results obtained from fitting the ZF-$\mu$SR signals of YBa$_2$Cu$_3$O$_{6.985}$ from 
Ref.~\onlinecite{Sonier:02} with $\Delta$ as a $T$-independent fit parameter. The features below 100~K are enhanced in the
higher doped sample and as shown in Fig.~\ref{fig2}(b) are well correlated with the changes in the
the Cu(2) NQR linewidth measured in YBa$_2$Cu$_3$O$_7$ (Ref.~\onlinecite{Grevin:00}). The small broad peak argued to
occur at $T^* \! \sim \! 77$~K by Zhang {\it et al.} is in fact not a peak, rather there is a dip in $\lambda_{\rm ZF}$ near
60~K that occurs on an otherwise increased or increasing $\lambda_{\rm ZF}$ below 100~K. This is evident from the NQR data, which show
no feature near 77~K, but instead exhibit a local maximum near 60~K. As explained in Ref.~\onlinecite{Sonier:02},
other kinds of experiments provide evidence for an unbuckling of the CuO$_2$ layers near 60~K.
Even in the absence of CDW order,
such a structural change modifies the nuclear dipole contribution to the ZF-$\mu$SR signal by changing the distance
between the $\mu^+$ and the host nuclei. Consequently, the effect is a modificaton of the
relaxation rate near 60~K, which has been observed over a wide doping range from underdoped YBa$_2$Cu$_3$O$_{6.50}$ 
through to overdoped (Ca-doped) samples.\cite{Sonier:02,Sonier:09} 

\begin{figure}
\centering
\includegraphics[width=8.5cm]{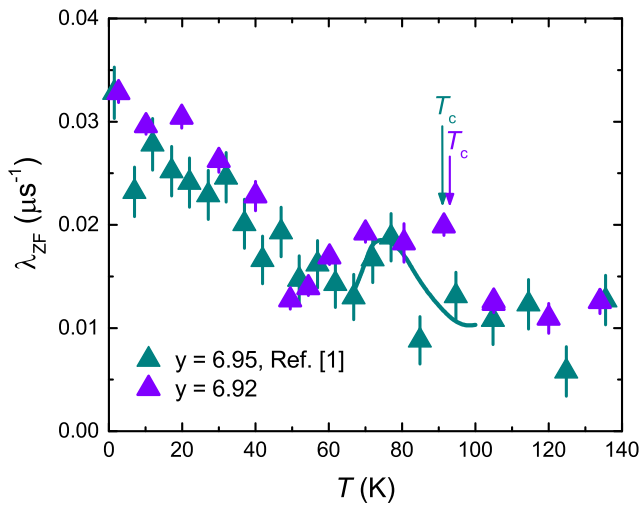}
\caption{(Color online) Results of fits to Eq.~(\ref{eqn:GE}) assuming $\Delta$ is $T$-independent. 
Comparison of the temperature dependence of the exponential relaxation rate $\lambda_{\rm ZF}$
in YBa$_2$Cu$_3$O$_{6.95}$ from Ref.~\onlinecite{Zhang:17} to $\lambda_{\rm ZF}$ obtained from fits of the 
YBa$_2$Cu$_3$O$_{6.92}$ ZF-$\mu$SR spectra reported in Ref.~\onlinecite{Sonier:02}.}
\label{fig3}
\end{figure}

A comparison of the results of similar $T$-independent $\Delta$ fits of 
the ZF-$\mu$SR spectra for YBa$_2$Cu$_3$O$_{6.92}$ from Ref.~\onlinecite{Sonier:02} to the YBa$_2$Cu$_3$O$_{6.95}$ data of 
Zhang {\it et al.} is shown in Fig.~\ref{fig3}. The good agreement between these independent measurements
indicates that sample quality and/or experimental factors are not relevant to the interpretation of the ZF relaxation data.
Again, the slight maximum of $\lambda_{\rm ZF}$ near 77~K is not due to critical slowing down of magnetic fluctuations,
but rather an artifact of assuming the nuclear dipole contributrion to the ZF-$\mu$SR signal is described by
a $T$-independent static Gaussian KT function. As for the case of YBa$_2$Cu$_3$O$_{6.985}$, better fits are achieved
with $\Delta$ free to vary with temperature.     
      
Zhang {\it et al.} also claim to have measured a small peak in the ZF relaxation rate of YBa$_2$Cu$_3$O$_{6.72}$ and a small
broad peak in the LF relaxation rate of YBa$_2$Cu$_3$O$_{6.77}$ (in weak LF) near $T^*$ at 210~K and 160~K, respectively.
At 160~K the muon is on the verge of diffusing and is clearly diffusing at 210~K, as established in measurements of 
YBa$_2$Cu$_3$O$_{6.67}$ in Ref.~\onlinecite{Sonier:02}. 
The width of the internal field distribution associated with the nuclear dipoles in YBa$_2$Cu$_3$O$_y$ is sufficiently 
narrow that the muon spin polarization in ZF decreases over the full time range of the recorded ZF-$\mu$SR spectra ({\it i.e.}
up to 14~$\mu$s). Slow muon diffusion reduces the polarization decay at later times, but with increased $\mu^+$ hopping rate $\nu$
the ZF-$\mu$SR signal evolves into an exponential function with a relaxation rate that decreases with increasing $\nu$.\cite{Schenck} 
In other words, the effect of muon diffusion on the ZF-$\mu$SR signal of YBa$_2$Cu$_3$O$_y$ is a reduction of the relaxation rate.
As shown in Fig.~2(a) of Ref.~\onlinecite{Zhang:17}, $\lambda_{\rm ZF}$ in YBa$_2$Cu$_3$O$_{6.72}$ decreases above 
$T \! \sim \! 160$~K. A decrease of $\lambda_{\rm ZF}$ above 160~K
is also evident in Fig.~\ref{fig2}(b) here, where the Cu(2) NQR linewidth is $T$-independent.
The small maximum in $\lambda_{\rm ZF}$ observed in YBa$_2$Cu$_3$O$_{6.72}$ near 210~K could
originate from the mobile $\mu^+$ reaching and becoming trapped by defects during its short lifetime.
The important point here is that the internal fields sensed by the muon have a time dependence above 
$T \! \sim \! 160$~K due to muon diffusion, which renders the $\mu^+$ an ineffective probe of
critical magnetic fluctuations at higher temperatures. 
 
When a static LF much greater than the nuclear dipole field distribution is applied, the muon spin decouples 
from the nuclear dipoles such that the nuclear contribution to the relaxation vanishes. However, when the muon is 
diffusing it may experience fluctuating field components transverse to the applied static LF that cause an increase 
in the LF relaxation rate. The reported maximum in $\lambda_{\rm LF}$ near 160~K in YBa$_2$Cu$_3$O$_{6.77}$ 
(Fig.~2(b) in Ref.~\onlinecite{Zhang:17}) may then result from the onset of muon diffusion and subsequent trapping
by defects at higher temperatures. It is important though to point out that the peaks claimed in the relaxation 
rates of the underdoped samples are extremely small ($\ll 0.01$~$\mu$s$^{-1}$), near the reliable detection limit of 
the method and on the order of statistical jumps in the data. Hence their very existence is questionable.

Finally, from LF-$\mu$SR measurements just above $T_c$ in fields $0.002 \! \leq \! B_{\rm LF} \! \leq \! 0.35$~T, Zhang {\it et al.} conclude that there are fluctuating magnetic
fields in the pseudogap regime with fluctuation rates as expected for ordered loop currents.
The correlation times and rms local field sensed by the muon are determined by fitting LF scans just above $T_c$
to the Redfield formula (see Fig.~1 in Ref.~\onlinecite{Zhang:17}). Yet the presented $\lambda_{\rm LF}$ versus $B_{\rm LF}$ 
data clearly deviate from the Redfield formula. Consequently, the quantitative
information obtained from these fits is invalid. While there does appear to be a gradual
reduction of $\lambda_{\rm LF}$ with increasing $B_{\rm LF}$ for two of the three data sets, the values and variation of 
$\lambda_{\rm LF}$ are so small (on the order of 10$^{-3}$~$\mu$s$^{-1}$) that extrinsic effects cannot be ruled out.
For example, the incoming positive muons experience magnetic field components perpendicular to their momentum
from the fringe fields at the end of the magnet used to generate $B_{\rm LF}$. This modifies the muon beam focusing.
A gradual reduction of $\lambda_{\rm LF}$ with increasing $B_{\rm LF}$ may then originate from a small increase in
the fraction of muons that miss the sample. The size of this effect will depend on the sample size and
radial position of the sample in the muon beam. Because the LF relaxation rates are below the typical
reliabe detection limit, the control experiments performed by Zhang {\it et al.}
on pure Ag can only rule out this scenario if the precise sample size, shape and position are replicated.

\end{document}